\documentclass[epsfig,12pt]{article}
\usepackage{epsfig}
\usepackage{graphicx}
\usepackage{array}
\usepackage{caption}
\usepackage{subcaption}
\usepackage{slashed}
\usepackage{amsmath}
\usepackage{braket}
\usepackage{verbatim}
\usepackage{amsfonts}

\newcommand{\beq}{\begin{equation}}   
\newcommand{\eeq}{\end{equation}}
\newcommand{\beqn}{\begin{eqnarray}}   
\newcommand{\eeqn}{\end{eqnarray}}

\newcommand{\bea}{\begin{eqnarray}}
\newcommand{\eea}{\end{eqnarray}}
\newcommand{\be}{\begin{equation}}
\newcommand{\ee}{\end{equation}}
\newcommand{\bead}{\begin{aligned}}
\newcommand{\eead}{\end{aligned}}

\newcommand{\nn}{\nonumber}

\newcommand{\gsim}{\lower.7ex\hbox{$
\;\stackrel{\textstyle>}{\sim}\;$}}
\newcommand{\lsim}{\lower.7ex\hbox{$
\;\stackrel{\textstyle<}{\sim}\;$}}
\begin{document}

\begin{titlepage}

\begin{flushright}
FTPI-MINN-15-14, UMN-TH-3428-15\\
% November 11, 2014
\end{flushright}

\vspace{0.4cm}

\begin{center}
{  \large \bf  't Hooft-Polyakov Monopoles with Non-Abelian Moduli}
\end{center}
\vspace{0.3cm}

\begin{center}
 {\large 
M. Shifman$^a$, G.Tallarita$^{b,c}$, and A. Yung$^{a,d}$}
\end {center}

\vspace{0.1mm}
 
\begin{center}

$^a${\em William I. Fine Theoretical Physics Institute, University of Minnesota,
Minneapolis, MN 55455, USA}\\[1mm]
$^b${\em Centro de Estudios Cient\'{i}ficos (CECs), Casilla 1469, Valdivia, Chile}
\\[1mm]
$^c${\em Departmento de F\'{i}sica, Universidad de Santiago de Chile, Casilla 307, Santiago, Chile}
\\[1mm]
$^{d}${\em National Recearch Center ``Kurchatov Institute'', 
Petersburg Nuclear Physics Institute, Gatchina, 
St. Petersburg 188300, Russia}

\end {center}

\vspace{0.5cm}

\begin{center}
{\large\bf Abstract}
\end{center}  

We extend the Georgi-Glashow model of the t'Hooft-Polyakov monopoles to include additional 
collective coordinates Ð  ``orientational isospin moduli." The low-energy theory of these solitonic solutions can be interpreted as dyons with isospin.

\hspace{0.3cm}

\end{titlepage}

%\tableofcontents
\section{Introduction}

Magnetic monopole is one of the most venerable constructions in theoretical physics. 
It dates back to Dirac \cite{Dirac}. Implementation of this construction in field theory is due to
't Hooft \cite{Hooft} and Polyakov \cite{Polyakov} (for  reviews and references see e.g. \cite{Weinberg,Rubakov,MAS}).
Remarkable effects were shown 
to be associated with the 't Hooft-Polyakov monopoles, e.g. the Callan-Rubakov effect of the baryon decay catalysis
\cite{A,B,C}. In the monopole field $SU(2)_{\rm gauge}$ doublet fermions
acquire integer spin (e.g. \cite{Rubakov}). This phenomenon is called ``spin from isospin."

In this paper we will consider an  extension of the  Georgi-Glashow model \cite{Georgi:1972cj}, in which a {\em global}
$O(3)$  (``isospin")  symmetry is present in the Lagrangian.\footnote{This latter isospin results from the global $O(3)$
symmetry, not to be confused with $SU(2)_{\rm gauge}$.}
 A simple model possessing this property and suitable for our purposes was suggested 
in \cite{Shifman:2012vv} (see also \cite{SYung}). Conceptually it was  inspired by Witten's cosmic strings \cite{W}.
Our task is to demonstrate that in this model the 't Hooft-Polyakov monopole acquires additional collective coordinates,
isospin moduli. The overall set of collective coordinates includes three coordinates of the monopole center,
a $U(1)$ phase coordinate which, upon quantization, corresponds to the electric charge and produces dyons out of the monopoles, and two extra collective coordinates associated with the $O(3)$ isospin. Quantization of the
isospin collective coordinates is straightforward, paralleling that of the spherical quantum top.

\section{The model}

Our starting point is the well-known Georgi-Glashow model \cite{Georgi:1972cj}. The gauge group is $SU(2)$ and the matter sector is described by a triplet real field $\phi^a$ (belonging to the adjoint representation). The Lagrangian of this model is
\be
\label{mono}
\mathcal{L}_{\rm GG} = -\frac{1}{4g^2} G^a_{\mu\nu}G^{\mu\nu,a}+\frac{1}{2} (D_\mu \phi^a)(D^\mu\phi^a)-\lambda\left(\phi^a\phi^a-v^2\right)^2,
\ee
where
\be
D_\mu\phi^a = \partial_\mu\phi^a +\varepsilon^{abc}A^b_\mu\phi^c,
\ee
is the covariant derivative in the  adjoint  representation, and
\be
G^a_{\mu\nu}=\partial_\mu A^a_\nu-\partial_\nu A^a_\mu+\varepsilon^{abc}A^b_\mu A^c_\nu
\ee
is the non-Abelian field strength. We use the following matrix notation for the fields
\be
\phi = \phi^a\frac{\tau_a}{2},
\ee
where $\tau_a$ denote the standard Pauli matrices. In the Lagrangian, $v$ is a parameter with dimensions of mass, and $\lambda$ is a dimensionless coupling constant. As is well known, this model supports the t'Hooft-Polyakov magnetic monopoles \cite{Hooft,Polyakov} that are topologically stable as a consequence of the symmetry breaking pattern
\be
SU(2)_{\rm gauge} \rightarrow U(1)_{\rm gauge}\,.
\label{sbp}
\ee
The breaking (\ref{sbp})  is enforced by the non-vanishing vacuum expectation value of the $\phi$ field, which can always be chosen aligned in the $3$-direction in $SU(2)_{\rm gauge}$,
\be
\phi^a_{vac} = v\delta^{3a}\,.
\ee
Two components of the gauge field, called $W^{\pm}$, acquire masses
\be
m_W = gv\,,
\ee
whilst the component aligned along the vacuum direction remains massless and plays the role of the $U(1)$ photon. The mapping between the group space and the coordinate space at infinity can be classified by the second
homotopy class,
\be
\pi_2\left(SU(2)/U(1)\right)=\mathbb{Z},
\ee
which guarantees the topological stability of the monopoles.   

Historically, the introduction of non-Abelian moduli on topological defects occurred in a rather advanced settings, mostly involving supersymmetric gauge theories (see \cite{AHDT,book,Gorsky}). In \cite{Shifman:2012vv}  a much simpler setup was suggested resulting in the occurrence of non-Abelian moduli. We will use this setup to study non-Abelian moduli on the world-line of the t'Hooft-Polyakov monopoles. To this end we will end a term $\mathcal{L}_{\chi}$ to the
Lagrangian (\ref{mono}),
\be
\mathcal{L} = \mathcal{L}_{\rm GG}+\mathcal{L}_{\chi}
\ee  
where 
\be
\mathcal{L}_{\chi}=\partial_\mu\chi^i\partial^\mu\chi^i-\gamma\left[(-\mu^2+\phi^a\phi^a)
\chi^i\chi^i+\beta(\chi^i\chi^i)^2\right].
\label{10}
\ee
Here $\chi^i$ is an $O(3)$ triplet uncharged under the gauge group. The potential term coupling $\chi^i$ to $\phi^a$ ensures that, if $\mu < v$, the $\chi^i$ field will develop in the monopole core and vanish in the vacuum. The mass of the $\chi$ field is
\be
m_\chi^2=\gamma(v^2-\mu^2).
\ee
The global $O(3)$ symmetry of (\ref{10}) is obvious.

Now, assuming the standard monopole ansatz for the fields $\phi$ and $A$
\be
\phi^a = v n^a H(r), \quad A_i^a = \epsilon^{aij}\frac{1}{r} n^j F(r),
\ee
with $n^a= x^a/r$, and adding a self-evident ansatz for $\chi^i$,
\be
\chi^i = \sqrt{\frac{\mu^2}{2\beta}}\;\chi(r)\;\left (
\begin{array}{l}
0\\0\\1
\end{array}
\right),
\ee
we can reduce 
the energy functional  to 
\bea\label{energy}
E &=& \frac{4\pi v}{g}\int_0^\infty d\rho\rho^2\bigg[\frac{(F')^2}{\rho^2}+\frac{(2F-F^2)^2}{2\rho^4}+\frac{H^2(1-F)^2}{\rho^2}+\frac{(H')^2}{2}\nn\\ [2mm]
&&+\frac{\lambda}{g^2}\left(H^2-1\right)^2+\frac{\tilde{\mu}^2}{2\beta}\left( (\chi')^2+\frac{\gamma}{g^2}\left[(-\tilde{\mu}^2+H^2)
\chi^2+\frac{\tilde{\mu}^2}{2}\chi^4\right]\right)\bigg],
\eea
in dimensionless units 
\be
\rho=gv|\vec{x}|\,.
\ee
Here $\tilde{\mu}$ is dimensionless,
\be
\tilde{\mu} = \mu/v.
\ee 
The energy minimization equations are
\bea\label{eom1}
\left(\rho^2 H'\right)'
&=&H\left[2(F-1)^2+\frac{4\lambda\rho^2}{g^2}\left(H^2-1\right)+\frac{\gamma\tilde{\mu}^2\rho^2}{g^2\beta}\chi^2\right],\\[2mm]
\label{eom2}
\rho^2 F''
&=&
(-1+F)\left((-2+F)F+\rho^2H^2\right),\\[2mm]
\label{eom3}
\left(\rho^2\chi'\right)'
&=&
\frac{\gamma\rho^2}{g^2}\chi\left(H^2+\tilde{\mu}^2\left(-1+\chi^2\right)\right).
\eea

\subsection{Vacuum structure}

In the vacuum, all derivatives of the profile functions must vanish. Clearly equation (\ref{eom2}) imposes the condition that $F=1$ in the vacuum. This leads to four branches of extrema of the energy, the first is
\be
H^I_{vac} = \chi^I_{vac}=0,
\ee
this is a maximum of the energy functional with $E^I_{max}/4\pi v=\lambda/g^3$. Then, in the vacuum II,
\be
H^{II}_{vac}=1,\quad \chi^{II}_{vac}=0,
\ee
for which the vacuum energy density vanishes $E^{II}_{vac}=0$. The third branch is
\be
H^{III}_{vac}=0,\quad \chi^{III}_{vac}=1,
\ee
for which
\be
\frac{E^{III}_{vac}}{4\pi v}=\frac{1}{g^3}\left(\lambda-\frac{\tilde{\mu}^4\gamma}{4\beta}\right).
\ee
Finally, the fourth branch is
\be
(H^{IV}_{vac})^2 = \frac{4\beta\lambda-\gamma\tilde{\mu}^2}{4\beta\lambda-\gamma},\quad (\chi^{IV}_{vac})^2=\frac{4\beta\lambda(1-\tilde{\mu}^2)}{\tilde{\mu}^2(\gamma-4\beta\lambda)}
\ee
for which
\be
\frac{E^{IV}_{vac}}{4\pi v}=\frac{\gamma\lambda}{g^3}\frac{(\tilde{\mu}^2-1)^2}{\gamma-4\beta\lambda}.
\ee
Hence, when $\gamma<4\beta\lambda$ this branch corresponds to the actual vacuum. The branches $II$ and $IV$ meet at $\tilde{\mu}=1$, or when $\mu=v$. When $\gamma > 4\beta\lambda$ and $\lambda > \frac{\tilde{\mu}^4\gamma}{4\beta}$ the vacuum $II$ is the actual vacuum.
\vspace{2mm}

Below we look for solutions in vacuum $II$.

\section{Numerical results}

Here we present the solutions of (\ref{eom1})-(\ref{eom3}) obtained by a second order central finite difference numerical procedure with the accuracy ${O}(10^{-4})$. To launch our numerical procedure we cut-off the radial direction at a large value of $\rho$, which we will call $R$. We set $R=100$.
As an example we choose the following values of the parameters:
\be
\label{26}
\gamma=4, \quad \lambda=0.34,\quad \tilde{\mu}=0.999, \quad\beta=2.94, \quad g=0.6\,.
\ee
Our boundary conditions are
\be
H(0)=F(0)=0, \quad \chi'(0)=0,
\ee
\be
H(R)=F(R)=1, \quad \chi(R)=0.
\ee
 The sought for solution is shown in Figure 1. 
The energy (monopole mass) $M_M$ of this solution is
\be
\frac{M_M}{4\pi v} = 2.14\,, \qquad \mbox{Monopole with the isospin}.
\ee
The mass $M_M$ of the standard t'Hooft-Polyakov monopole at $\lambda = 0.34$ and $g=0.6$ is 
\be
\frac{M_M}{4\pi v} = 2.33\,.
\ee
Thus,  the energy of the monopole with the nonvanishing $ \chi^i$ in the core is lower.

 \begin{figure}[ptb]
\centering
\includegraphics[width=0.8\linewidth]{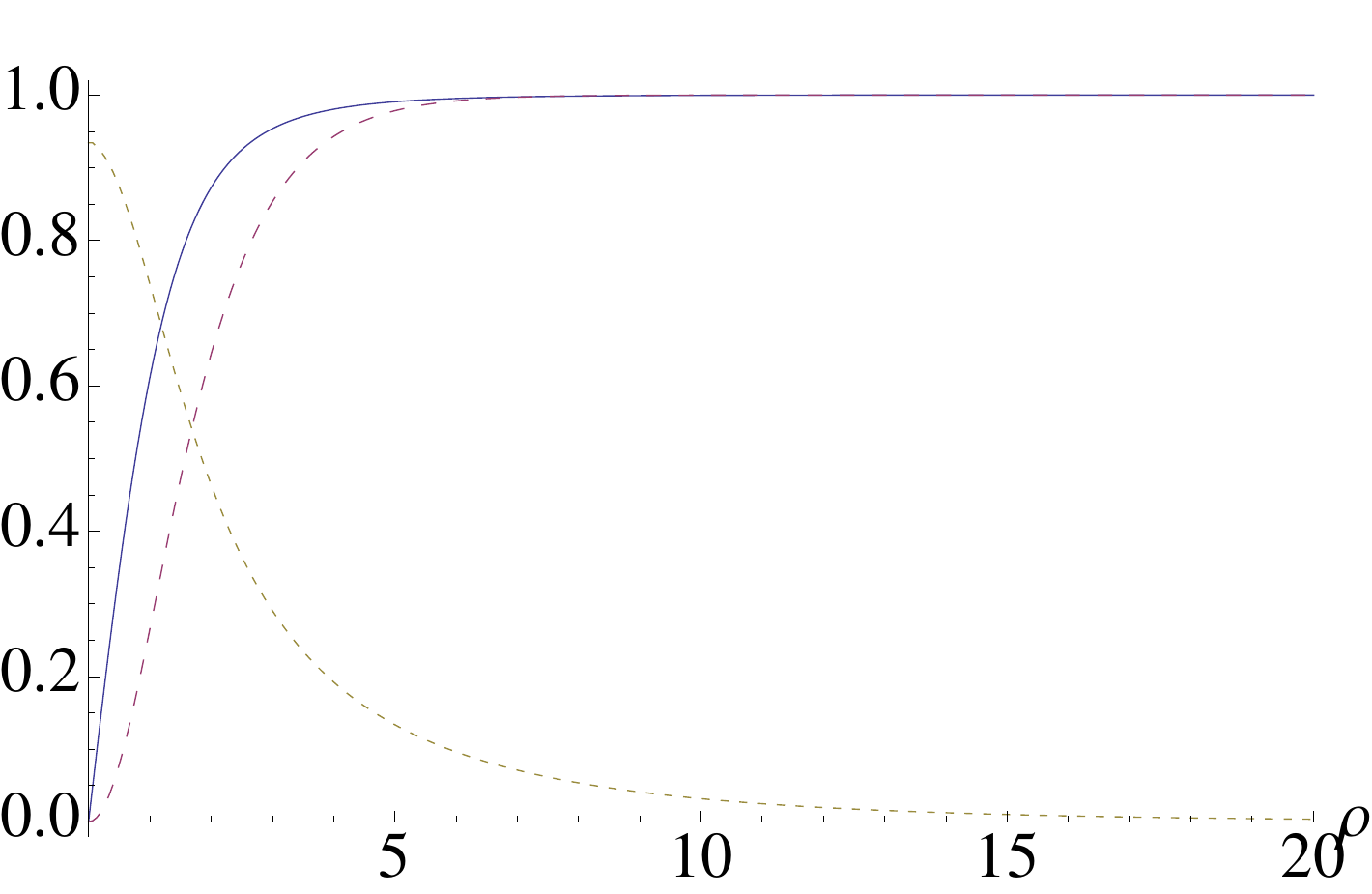} \caption{Non-Abelian monopole profiles for $\gamma=4$, $\lambda=0.34$, $\tilde{\mu}=0.999$, $\beta=2.94$ and $g=0.6$. The solid curve represents $H$, medium dashed $F$ and thin dashed $\chi$.}%
\label{fig1}%
\end{figure}

\section{Stability}

In this section we can check that the solution found above with a nonvanishing $\chi$ in the core is stable.  The energy functional for the $\chi$ field is
\be
E_{\chi} = \frac{4\pi v}{g}\int_0^\infty d\rho\rho^2\left[(\chi')^2+\frac{\gamma}{g^2}\left((-\tilde{\mu}^2+H^2)
\chi^2+\frac{\tilde{\mu}^2}{2}\chi^4\right)\right].
\ee
Introducing
\be
\psi(\rho)=\rho\chi
\ee
we can derive a one-dimensional Schr\"odinger-like equation for the $\psi$ eigenfunctions,
\be
-\psi''+\frac{\gamma}{g^2\rho^2}\left((-\tilde{\mu}^2+H_1^2)+3\tilde{\mu}^2\chi_1^2\right)\psi=\epsilon\psi
\ee
where $H_1$ and $\chi_1$ are the field profiles shown in Figure 1. Using the same parameters as in (\ref{26}) we find the lowest-lying mode at $\epsilon\approx 0.00105$. The positivity of $\epsilon$ demonstrates the  stability of the numerical solution with a non-vanishing $\chi$ field in the core. 

Since the solution with a non-zero $\chi$ in the core has lower energy than the vanishing-$\chi$ solution we must also investigate (meta)stability of the vanishing-$\chi$ solution. Then, following the argument above, we must solve 
\be
-\psi''+\frac{\gamma}{g^2\rho^2}(-\tilde{\mu}^2+H_1^2)\psi=\epsilon\psi\,.
\ee
We find that the lowest lying mode is at $\epsilon\approx 0.00103$ thus confirming the classical stability of the $\chi=0$ solution.

Thus, we observe two classical solutions: one -- the standard 't Hooft-Polyakov monopole is a local minimum of the energy functional, while the solution with the nonvanishing $\chi$ in the core represents the global minimum.

\section{Quantization of the collective coordinates}

Quantization of the standard t'Hooft-Polyakov monopole involves four moduli: three translations of the monopole center plus a rotation around the vacuum direction in the $SU(2)_{\rm gauge}$ space. In the adiabatic approximation one finds the  Lagrangian  
\be
\mathcal{L}_{QM} = - M_M+\frac{M_M}{2}(\dot{\vec{x_0}})^2
+\frac{1}{2}\frac{M_M}{m^2_W}\dot{\alpha}^2
\ee
where $M_M$ is the mass of the monopole, the vector $\vec{x_0}$ represents the monopole center, whilst $\alpha$ is the collective coordinate  related to the remaining $U(1)_{\rm gauge}$ invariance.  The full quantum mechanical Hamiltonian for these moduli is
\be
\widehat{H}_0 = M_M +\frac{\vec{p}\;^2}{2M_M}+\frac{1}{2}\frac{m_W^2}{M_M}\pi^2_\alpha,
\label{36}
\ee
where
\be
\pi_\alpha = \frac{M_M}{m^2_W}\dot{\alpha},
\ee
is the canonical momentum conjugated to the angular variable $\alpha$. 

To obtain the orientational moduli we parametrize the $\chi$ field as follows 
\be\label{para}
\chi^i = \sqrt{\frac{\mu^2}{2\beta}} \chi(\rho)S^i(t)
\ee
where $S^i$ is a unit vector which depends on time. Substituting (\ref{para}) in (\ref{energy}) we obtain the low-energy action for the orientational moduli, 
\be\label{lagg}
\mathcal{S}_{0} = \frac{I_1}{2} \int dt\;  \dot{S}^i\dot{S}^i , \quad S^iS^i=1,
\ee
where
\be
\frac{I_1 v }{4\pi}= \frac{\tilde{\mu}^2}{g^3\beta}\int_0^\infty d\rho\,\rho^2\chi^2 =7.81,
\ee
and the overdot denotes a derivative with respect to time.  The above action can be easily quantized as follows. 

The action (\ref{lagg}) describes the motion of a rigid rotating body, symmetric quantum top, at a fixed spherical radius $r_0$ (in this case $r_0=1$).   Its quantization can be carried out in a standard way (see. e.g. \cite{slm}), if we parametrize $S^i $ in terms of polar and azimuthal angles,
\be
S^1= \cos\theta\,,\quad S^2= \sin\theta\,\cos\varphi \,,\quad S^3= \sin\theta\,\sin\varphi\,.
\ee
Upon quantization we get the following symmetric top Hamiltonian:
\beq
\widehat{H} = -\frac{1}{2I_1}\left[ \frac{1}{\sin{\theta}} \frac{\partial}{\partial\theta}
\left(\sin{\theta}\frac{\partial}{\partial\theta}\right) 
+\frac{1}{\sin^{2}\theta}\frac{\partial^{2}}{\partial\varphi^{2}}
\right].
\eeq
Its eigenvalues are labelled by a non-negative integer
$s$ and reduce to
\be
E_{s}=\frac{1}{2I_{1}} s\left(  s+1\right)  \,.
\ee
They have degeneracy $2s+1$.
The eigenfunctions are the standard spherical harmonics $Y_{s}\left(
\theta,\varphi\right)  $. 

Including the conventional 't Hooft-Polyakov moduli from Eq. (\ref{36}) we arrive at the monopole mass formula 
\beq
M = M_{M } +  \frac{ m_W^2 \,k^2}{2M_{M }}\ +  \frac{1}{2I_1}\,s(s+1),
\eeq
%\vec{p}^{\,2} +
where $k$ and $s$ are integers. The $k^2$ term here corresponds to  dyonic
excitations associated with non-zero electric charge, while the last term is
associated with non-zero global isospin. The $I_1$ parameter plays the role of the moment of 
inertia in the isospace.
With our illustrative choice of parameters $I_1^{-1} \sim (1/60) m_W$ while $m_W/M_M \sim 1/40.$

\section{Conclusions}

In this model we completed the program of constructing the simplest topological defects
with non-Abelian moduli. 
We extended the Georgi-Glashow model of the t'Hooft-Polyakov monopoles to include extra global symmetry $O(3)$
(we referred to it as isospin) 
and extra collective coordinates Ð  ``orientational isospin moduli." The adiabatic quantization of these solitonic solutions
was carried out. The result  can be interpreted as a dyon with isospin.

 \section*{Acknowledgments}

This work  is supported in part by DOE grant DE-SC0011842 and Fondecyt grant No. 3140122. 
The work of A.Y. was  supported by William I. Fine Theoretical Physics Institute  of the  University of Minnesota,
by Russian Foundation for Basic Research under Grant No. 13-02-00042a and by Russian State Grant for
Scientific Schools RSGSS-657512010.2. The work of A.Y. was supported by Russian Scientific Foundation 
under Grant No. 14-22-00281. The Centro de Estudios Cient\'{i}ficos (CECS) is funded by the Chilean
Government through the Centers of Excellence Base Financing Program of Conicyt.

\end{document}